\documentclass[prl,aps,floats,superscriptaddress,floatfix,twocolumn]{revtex4}
\usepackage{amssymb,amsmath}
\usepackage{graphicx}
\usepackage{xcolor}
\usepackage{soul}
\usepackage{dcolumn}
\usepackage[utf8]{inputenc}
\usepackage{hyperref}

\usepackage{bm}
\usepackage[normalem]{ulem}
 \hypersetup{
    colorlinks=true,
    citecolor=blue,
    }
 \usepackage{amsmath}
 \usepackage{diagbox}
\usepackage{array}
\usepackage{tikz}

\newcommand{\magpentagon}{%
\tikz[baseline=-0.5ex, rotate=90]\draw[fill=magenta!50, draw=black!50]
(0:0.1cm) -- (72:0.1cm) -- (144:0.1cm) -- (216:0.1cm) -- (288:0.1cm) -- cycle;}

\newcommand{\graydiamond}{%
\tikz[baseline=-0.5ex]\draw[fill=gray!90, draw=black!40]
(0,0.10cm) -- (0.10cm,0) -- (0,-0.10cm) -- (-0.10cm,0) -- cycle;}

\begin{document}

\title{Universality of shocks in conserved
driven single-file motions with bottlenecks}
\author{Sourav Pal}\email{isourav81@gmail.com}
\affiliation{Theory Division, Saha Institute of
Nuclear Physics,  {a} CI of Homi Bhabha National Institute, 1/AF Bidhannagar, Calcutta 700064, West Bengal, India}

\author{Abhik Basu}\email{abhik.123@gmail.com, abhik.basu@saha.ac.in}
\affiliation{Theory Division, Saha Institute of
Nuclear Physics, {a} CI of Homi Bhabha National Institute, 1/AF Bidhannagar, Calcutta 700064, West Bengal, India}

\begin{abstract}
 Driven single-file motion,  in which particles move unidirectionally along one-dimensional channels,  sets the paradigm for wide variety of one-dimensional directed movements, ranging from intracellular transport and urban traffic to ant trails and controlled robot swarms. Motivated by the phenomenologies of these systems in closed geometries, regulated by number conservation and bottlenecks, we explore the domain walls (DWs) or shocks in a conceptual one-dimensional cellular automaton with a fixed particle number and a bottleneck.  For high entry and exit rates of the cellular automaton, and with sufficiently large particle numbers, the DWs formed are independent of the associated rate parameters, revealing a {\em hitherto unknown universality} in their {\em shapes}, which are however enclosed by nonuniversal boundary layers.  In contrast, the DWs do depend upon these parameters, if small, and hence have nonuniversal shapes, but without boundary layers. Nonuniversal delocalized DWs can be formed by additional tuning of the control parameters. Our predictions on the DWs are testable in model experiments.
\end{abstract}

\maketitle

Driven single-file motion (DSFM) is the cornerstone of effective one-dimensional (1D) transport without overtaking in physical systems of diverse origin. This includes intracellular transport, e.g., motion of molecular motors along eukaryotic cells~\cite{cell}, ribosome translocation along messenger RNA (mRNA) strands~\cite{mrna22,mrna33,mrna44}, vehicular or pedestrian traffic along closed network of
roads~\cite{traffic1,traffic2,bus-route}, ecological examples like ant trails~\cite{ant1,ant2} and also 1D motion of robot swarms~\cite{robot-2024}.   In closed systems of DSFM connected to a particle storage or reservoir,
regulated by the interplay of the overall fixed availability of particles, DSFM-reservoir couplings and bottlenecks, the  stationary density profiles and their degree of {\em universality} vis-\`a-vis the control parameters in the system remains a theoretically and phenomenologically paramount issue.

In this Letter, motivated by the above generic issues in DSFM, we propose and study a conceptual 1D cellular automaton having a fixed number of available particles with only excluded volume interactions, and a bottleneck, based on totally asymmetric simple exclusion process (TASEP)~\cite{krug,derrida1,derrida2}. We use it to explore universal shapes of the  domain wall (DW) in the nonequilibrium steady states of the model.

 Originally proposed to study protein synthesis in eukaryotic cells~\cite{macdonald}, TASEP has subsequently emerged as a paradigmatic 1D model for nonequilibrium phase transitions in open boundary ~\cite{krug,derrida1,derrida2,derrida3,kolomeisky-1998,blythe,chou,atri1,atri2,sm-ab-tasep}, and closed geometries~\cite{lebo,lebo1,mustansir,ha-nijs-2002,erwin-defect,hinsch,basu-mohanty,corstin-2012,rakesh1,niladri1,r-basu,niladri-tirtha,rakesh2,soh,tirtha-qxtasep,parna-anjan,atri3}. 
TASEP models with quenched~\cite{lebo,motor,kolomeisky-1998,tripathy-barma97} or dynamical disorder~\cite{bhatia23,waclaw19} have been widely investigated. Notable is the \textit{slow-bond problem}~\cite{lebo,ha-timonen-nijs,soh,schmidt15,soh18,corstin-2012}, in which a single localized defect (or a slow bond) on a periodic lattice induces a phase segregation into coexisting low- and high-density regions, separated by a sharp stationary domain wall whose profiles depend only on the defect strength and the global particle density. 
Subsequent studies revealed that this local inhomogeneity can trigger a dynamic queuing phase transition beyond a critical defect strength~\cite{ha-timonen-nijs,schmidt15}. The interplay between jamming and condensation phenomena in modified TASEP models has also been explored in~\cite{soh18}. More recently, considerations of the finite availability of ribosomes in the protein synthesis in cells~\cite{cell} have led to studies of  TASEP lanes connected to particle reservoirs, revealing the role of global particle number conservation (PNC)~\cite{reser1,reser2,reser3,klumpp1,klumpp2,klumpp3,ciandrini,dauloudet,astik-parna,sourav1,sourav2,sourav3,astik-erwin}. Our work complements and extends the existing studies by exploring how the interplay between conservation laws, finite resources and local inhomogeneity ultimately determine the steady states and DWs.

\begin{figure}[!h]
 \centerline{
 \includegraphics[width=0.9\linewidth]{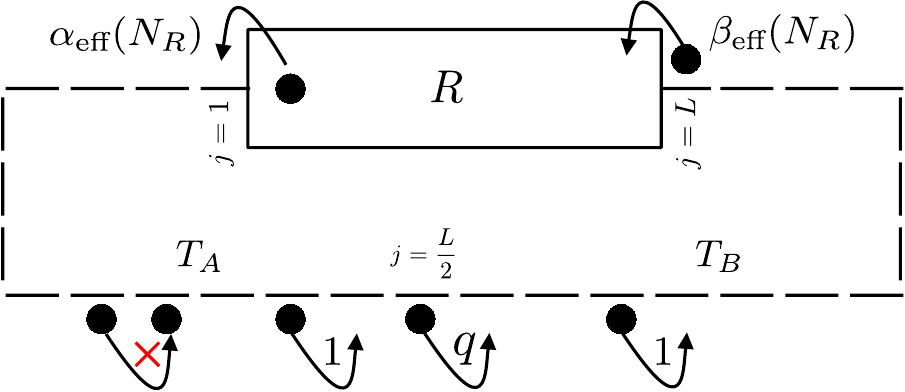}}
 \caption{Schematic model diagram. A 1D lattice executing TASEP  with hopping rate  $q<1$ at a point defect and unity elsewhere is connected to a particle reservoir $R$ at both ends. Effective entry and exit rates, $\alpha_\text{eff}$ and $\beta_\text{eff}$ respectively, depend explicitly on reservoir population $N_{R}$. See text.}
 \label{model-diagram}
 \end{figure}

Our model consists of a TASEP on a 1D lattice with $L$ sites ($j=1,2,...,L$), each having maximum occupancy 1, with unit hopping rate, except at the defect site at $j=L/2$, where hopping rate $q<1$; see Fig.~\ref{model-diagram}. Here, the effective entry rate $\alpha_\text{eff}$ and exit rate $\beta_\text{eff}$ are
\begin{align}
  \alpha_{\text{eff}}(N_{R}) &= \alpha f(N_{R}), \label{alpha-eff} \\
  \beta_{\text{eff}}(N_{R})  &= \beta (1 - f(N_{R})). \label{beta-eff} 
\end{align}
  A monotonically rising function, $f(N_R)$, of reservoir population $N_{R}$ describes the TASEP-reservoir couplings. It is physically reasonable to assume $\alpha_\text{eff}$ increasing with $N_R$ while $\beta_\text{eff}$ decreases~\cite{reser1,astik-parna,sourav1}. For simplicity, we choose $f(N_{R}) = N_{R}/N^{*} \in [0,1]$~\cite{astik-parna,sourav1,sourav2,sourav3,astik-erwin}, $N^{*}$ is a normalization factor and $\alpha,\,\beta \ge 0$; see also Ref.~\cite{seppa}. We define filling factor $\mu \equiv N_{0}/L$, where $N_0\equiv N_{R}+\sum_{j=1}^{L}n_{j}$ is the total particle number, a constant of motion, with {$n_{j} = 0,1$,} the occupation number of site $j$. The stationary states are parametrized by $\alpha,\,\beta,\,\mu,\,q,\,f$. Parameters $\alpha,\,\beta$ control the mutual interplay between the ``supply'' and ``demand'' in finite-resources systems, and model the DSFM-reservoir couplings, e.g., in mRNA translation~\cite{supply-demand}.
 
 As specific examples, we consider (i) Case I: $N^*=L$ and (ii) Case II: $N^*=N_0$. Since $\beta_\text{eff} > 0$, and using PNC, we must have (i) $N_0 \leq 2L$ or $0 \leq \mu \leq 2$ for $N^* = L$, corresponding to a finite carrying capacity of the reservoir for a given size $L$ of the TASEP lane; (ii) no such constraint on $N_0$ appears when $N^* = N_0$, allowing the model to accommodate any number of particles, i.e., $0 \leq \mu < \infty$, implying unrestricted carrying capacity of the reservoir. See also Refs.~\cite{sourav1,sourav2} in this context. Only for $N^* = L$ (but not for $N^* = N_0$), our model admits the particle-hole symmetry: $\alpha \leftrightarrow \beta$ and $\mu \leftrightarrow 2 - \mu$; see Supplemental Material (SM)~\cite{sm} for details.

  Generically the TASEP lane in the present model can be in spatially homogeneous or inhomogeneous steady states.  {The former class consists of the low-density (LD) and high-density (HD) phases, with bulk densities lesser and greater than $1/2$, respectively.} The inhomogeneous steady states can be a single localized DW (LDW), in which the stationary density is macroscopically nonuniform, which is static in time, or a pair of delocalized DWs (DDWs), in which density shocks are {\em moving} along the TASEP lane. There is however no maximal current (MC) phase (with bulk density 1/2), unlike in other models for open TASEP~\cite{blythe,kolomeisky-TASEP-phase} or TASEPs with finite resources~\cite{reser1,reser2,reser3,astik-parna,sourav1,astik-erwin,sourav2,jindal-arvind,arvind1,sourav3,kavcic2025}. 
 
  We show that 
 for sufficiently large  $\alpha,\,\beta$, and $\mu$, the defect in the TASEP lane can induce DWs that have {\em universal shapes},  independent of $\alpha,\,\beta,\,\mu$, and the form of $f$, {\em always} localized at the midpoint of the TASEP lane, with their heights depending {\em solely} on $q$. We call this DW a {\em universal} DW (UDW). Any measurement of a UDW tells us {\em nothing} about $\alpha,\,\beta,\,\mu,\,f$, and is reminiscent of universal critical phenomena~\cite{chaikin}. Rather intriguingly, such a UDW is necessarily associated with a pair of \textit{nonuniversal} boundary layers (BLs) depending on model parameters at the two ends of the TASEP lane, in contrast to usual TASEPs with open boundaries~\cite{blythe} or with finite resources~\cite{reser1,reser2,reser3,astik-parna,sourav1,astik-erwin,sourav2}.  For other choices of $\alpha,\,\beta$, conventional localized DWs without BLs emerge, induced either by the defect or the reservoir, which depend on various model parameters and hence are {\em nonuniversal}. Lastly, a pair of  DDWs with {\em nonuniversal} profiles appear for $\alpha,\,\beta$ on a curve in the $\alpha$-$\beta$ plane, determined by the {\em competition} between the reservoir and the defect. Identification of the UDW phase, which replaces the maximal current (MC) phase, and the role of {\em two nonuniversal} BLs in UDW formation are the principal results in this Letter.  As discussed below, these results can be explained by considering the TASEP lane to consist of {\em two} sub-lanes connected cyclically at the reservoir and point defect site, together with particle and current conservation.

\begin{figure*}[ht]
\includegraphics[width=\textwidth]{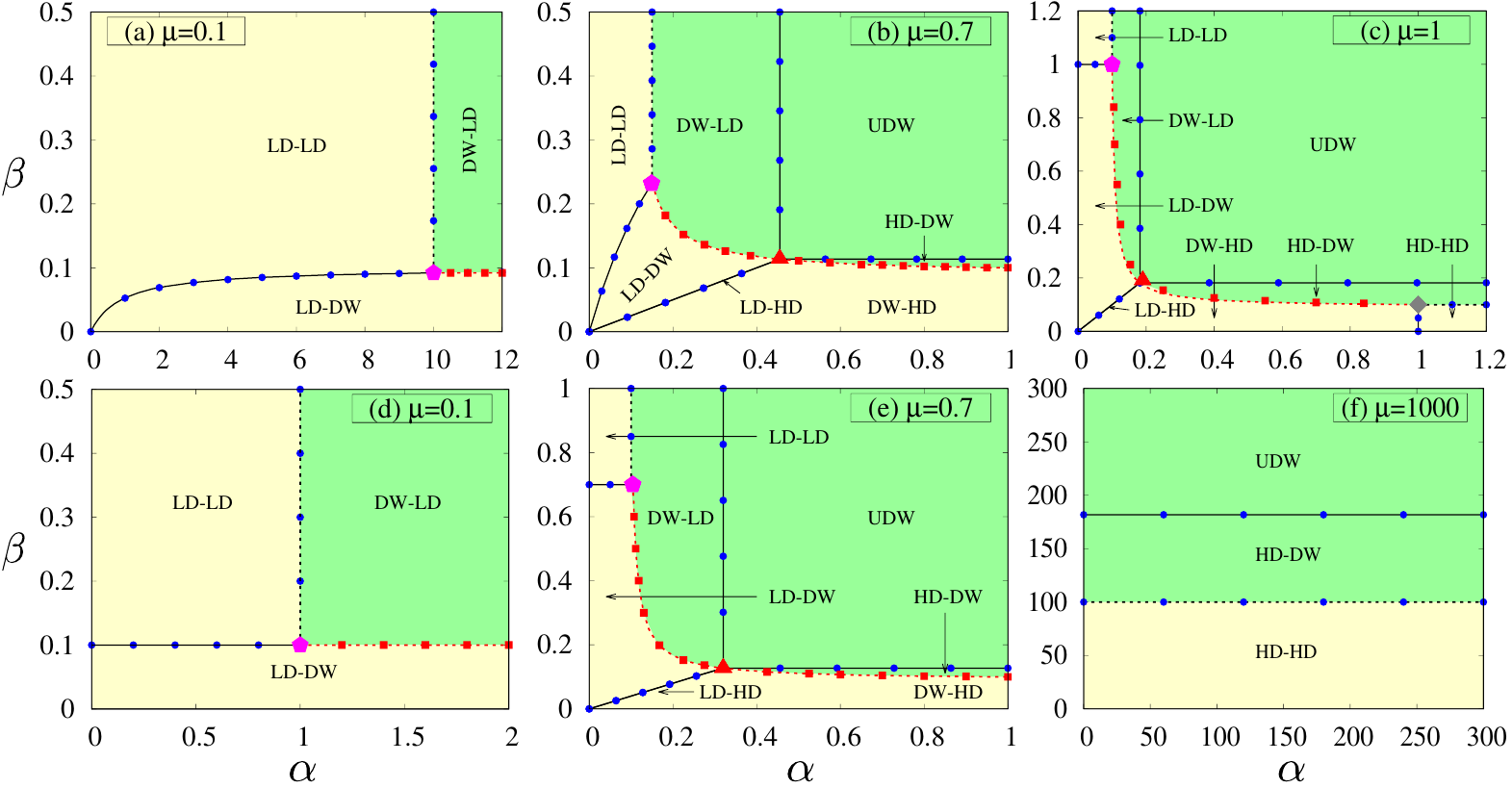}
\caption{Phase diagrams in the $\alpha$-$\beta$ plane at $q=0.1$ for representative values of $\mu$: $N^{*}=L$ [upper panel~(a–c)] and $N^{*}=N_{0}$ [lower panel~(d–f)]. Reservoir-controlled (defect-controlled) regions are shown in yellow (green). In total, eight distinct phases are present. Universal domain walls associated with nonuniversal boundary layers at the entry and exit ends form in the UDW phase. Delocalized DWs in $T_A$ and $T_B$ occur along the red dashed curve~(\ref{DDW-surface}) [panels~(a–e)], terminating at the points indicated by \protect\magpentagon\ (magenta pentagon) and \protect\graydiamond\ (gray diamond); see Table~I and Table~II of~\cite{sm} for their coordinates and existence ranges. Complete delocalization occurs only at $(\tilde\alpha,\tilde\beta)$ (see Eq.~(\ref{ab-tilde}) for coordinate), while partial delocalization occurs elsewhere. The point $(\tilde\alpha,\tilde\beta)$, marked by a red triangle, exists for $1/2<\mu<3/2$ when $N^{*}=L$ and for $1/2<\mu<\infty$ when $N^{*}=N_{0}$ (not shown for $\mu=1000$ in the $N^{*}=N_{0}$ case). Localized DWs are observed in the remaining DW regions. MFT predictions (solid and dashed curves) show good agreement with MCS data (symbols). See text.} 
\label{pd}
\end{figure*}

 For very low values of $\mu$, the bulk of the TASEP lane remains in the low-density (LD) phase. As $\mu$ rises, it moves to DW phases, controlled either by the defect or the reservoir, eventually reaching the high-density (HD) phase at large $\mu$; see movies~1 and~2 in SM~\cite{sm}. See the phase diagrams in Fig.~\ref{pd} in the $\alpha$-$\beta$ plane for various $\mu$-values with $q=0.1$, for both $N^{*} = L,\,N_{0}$, which differ significantly from that of an open TASEP~\cite{blythe}.

 Partly related models and studies appear in Ref.~\cite{brackley}, which investigates steady states with resource-dependent hopping rates and a slow site, and in Ref.~\cite{cook-slow}, which focuses on DWs. With $N^*=N_0$ and $q=1$, our model reduces to that studied in Ref.~\cite{astik-parna}.

 To calculate the phases and phase diagrams, we employ mean-field theory (MFT)~\cite{blythe} for the TASEP lane by modeling it as two segments, \(T_A\) and \(T_B\), connected at the defect site \(j = L/2\), each with their effective entry and exit rates. The MFT results are validated through extensive Monte Carlo simulations (MCS). 
Introducing the quasi-continuous coordinate \(x \equiv j/L\) in the thermodynamic limit (TL), \(L \to \infty\), with \(0 \leq x \leq 1\), we define the local density in segment $T_{m}$ by \(\rho_m(x) \equiv \langle \rho_j^m \rangle\), where $\rho_j^m$ is the occupation at site $j$, \(m = A,\,B\) and \(\langle \cdot \rangle\) indicates temporal averaging in the steady state.

With \((\alpha_\text{eff}, \beta_{A})\) as the effective entry and exit rates for  \(T_{A}\), and \((\alpha_{B}, \beta_\text{eff})\) as those for  \(T_{B}\), stationary current conservation at \(j = L/2\) and \(j = (L/2) + 1\) gives
\begin{eqnarray}
  &&\beta_{A}=q(1-\rho_{(L/2)+1}),\;\; 
  \alpha_{B}=q\rho_{L/2}. \label{beta-alpha-AB}
 \end{eqnarray}
 Additionally, current through the defect site reads
 \begin{equation}
  J_d=q\rho_{L/2}(1-\rho_{(L/2)+1})={\alpha_{B}\beta_{A}}/{q}. \label{current-through-defect}
 \end{equation}
  Denoting the steady-state mean density and current respectively by $\rho_{m}$ and $J_{m}$ for segment $T_{m}$, $m=A,\,B$, current conservation $J_{A}=J_{B}=J_d$ implies either $\rho_{A}=\rho_{B}$ or  $\rho_{A}+\rho_{B}=1$. In analogy with an open TASEP, the bulk density in the LD-LD phase is given by $\rho_A = \alpha_{\text{eff}}(N_R) = \alpha f(N_R)=\rho_B = \alpha_B$ (from current conservation), controlled by the reservoir.  Equation~(\ref{beta-alpha-AB}) together with PNC gives:
\begin{equation}
 \label{rholdld-L}
\rho_\text{LD-LD}=
\begin{cases}
    \mu \alpha/(1+\alpha),\;\;\; \; N^{*}=L, \\
    \mu \alpha/(\mu+\alpha),\;\;\; \; N^{*}=N_{0},
\end{cases}
\end{equation}
vanishing, unsurprisingly, for $\mu\rightarrow 0$. The defect imposes only a local jump of height $h=\alpha_\text{eff}(1-q)/q>0$ with a vanishing thickness in the TL behind it, as obtained by using current conservation in the steady-state~\cite{erwin-defect}.

The point defect can impose a DW behind it for larger $\mu$, connecting LD and HD  segments with densities $\rho_\text{LD}$ and $\rho_\text{HD}$ respectively. Stationary current conservation across the point defect gives~\cite{lebo,niladri1}
 \begin{eqnarray}
  && \rho_\text{LD}=q/(1+q)=1-\rho_\text{HD}. \label{rho-LD-HD-for-dw}
 \end{eqnarray}
 With $q\neq 1$, $\rho_\text{HD}>\rho_\text{LD}$. Thus,  {\em if} the defect controls the steady-state, there cannot be any MC phase.
 
 Qualitatively, as $N_{0}$ (or $\mu$) increases, $\rho_\text{LD-LD}$ rises [Eq.~(\ref{rholdld-L})], eventually reaching the defect-limited value $\rho_\text{LD-LD}\equiv\alpha_\text{eff}=q/(1+q)$. A DW is formed in $T_{A}$ at $x=1/2$, {marking the onset of the} DW-LD phase. With  $\Theta(x) = 1\,(0)$ for $x > (<)\,0$, for a DW at $x_{w_{1}}$ in $T_{A}$, $\rho_{A}(x)=\rho_\text{LD}+\Theta(x-x_{w_{1}})(\rho_\text{HD}-\rho_\text{LD})$, where $\rho_\text{LD}$ and $\rho_\text{HD}$ are as given in \eqref{rho-LD-HD-for-dw}. PNC gives
\begin{equation}
\label{xw-dwld}
x_{w_1} =
\begin{cases}
  \big[(1/2 - \mu)(1 + q) + q / \alpha \big] \big/ (1 - q), \\
  \big[(1/2 - \mu)(1 + q) + \mu q / \alpha \big] \big/ (1 - q).
\end{cases}
\end{equation}
for $N^*=L$ (top) and $N^*=N_0$ (bottom). Substituting $x_{w_{1}}=1/2$ in Eq.~(\ref{xw-dwld}) gives the boundary between LD-LD and DW-LD phases: $\mu \alpha/(1+\alpha) = q/(1+q)$ for $N^{*}=L$, and $\mu \alpha/(\mu+\alpha) = q/(1+q)$ for $N^{*}=N_{0}$ [see Figs.~\ref{pd}(a)-\ref{pd}(e)]. As  $\mu$ or $\alpha$ increases (with other parameters fixed), more particles enter the TASEP lane, pushing the DW leftwards, keeping $\alpha_\text{eff}$ unchanged, until $x_{w_1}=0$ at the onset of UDW phase. With $x_{w_{1}}=0$ in Eq.~(\ref{xw-dwld}), {$\alpha(\mu-1/2)=q/(1+q)$ for $N^{*}=L$, and $\alpha(1-1/2\mu)=q/(1+q)$ for $N^{*}=N_{0}$ [see Figs.~\ref{pd}(b, c, e)] are the boundary between DW-LD and UDW phases,  which do not depend on $\beta$.

For a DW at $x=1/2$ with HD and LD segments covering $T_A$ and $T_B$, respectively, BLs of vanishing thickness in TL form at $j=1$ and $j=L$. Current conservation gives the $\alpha,\,\beta,\,f,\,\mu,\,q$-dependent {\em nonuniversal} BL densities
\begin{eqnarray}
 &&\rho_{1} = \big[1+q/\{(1+q)\alpha(\mu-1/2)\}\big]^{-1}, \label{first-site-den} \\
&&\rho_{L} = \big[1+\{(1+q)\beta(3/2-\mu)\}/q\big]^{-1} \label{last-site-den}
\end{eqnarray}
for $N^{*}=L$; see
 SM~\cite{sm} for $N^{*}=N_{0}$.
As $\alpha$ and $\beta$ increase with $\mu$ and $q$ fixed, $\rho_1 \to 1$ and $\rho_L \to 0$, while the DW remains fixed at $x=1/2$, defining the UDW phase. For larger $\mu$, the DW however can move to $T_{B}$ crossing the entry-end $x=0$ of $T_A$  (same as the exit-end $x=1$ of $T_B$; see Fig.~\ref{model-diagram}), giving the HD-DW phase, and finally returns to $x=1/2$ giving the HD-HD phase. 
Current conservation in UDW and HD-DW phases gives $\rho_\text{LD}=q/(1+q)=1-\rho_\text{HD}$. As $q \to 1$, the DW height, $\rho_\text{HD} - \rho_\text{LD}$, gradually decreases vanishing at $q=1$, resulting in the MC phase.  The UDWs and BLs in this model can be rationalized considering that when a UDW is formed {\em behind} the local defect, i.e., in sublattice $T_A$, the stationary density should be controlled by its exit rate $q$. Furthermore, a UDW smoothly reduces to the MC phase density, which has two BLs on both sides, for $q\rightarrow 1$.  The UDW position is independent of {\em all} of $\alpha,\,\beta,\,q,\,\mu,\,f$, and the height depends {\em only} on $q$. Thus a UDW is controlled only by $q$. Hence in a renormalization group language, $q$ appears as the only relevant operator in UDW phase, with everything else {\em irrelevant}, reminiscent of universal critical phenomena~\cite{chaikin};  see also Ref.~\cite{tirtha-qxtasep} for universality in disordered TASEP. PNC gives the DW position 
\begin{equation}
\label{xw-hddw}
x_{w_{2}} = \left\{
\begin{aligned}
&\bigl[\,5/2 - \mu(1+q) + q (1/2 - 1/\beta)\bigr]\big/(1 - q), \\
&(\,3/2 - \mu q / \beta - q/2)/(1 - q),
\end{aligned}
\right.
\end{equation}
in HD-DW phase, for $N^{*}=L$ (top) and $N^{*}=N_{0}$ (bottom). Substituting $x_{w_{2}}=1$ {in Eq.~(\ref{xw-hddw})} gives the boundary between UDW and HD-DW phases: $\beta(3/2-\mu)=q/(1+q)$ for $N^{*}=L$, and $\beta=2 \mu q/(1+q)$ for $N^{*}=N_{0}$ [see Figs.~\ref{pd}(b, c, e, f)],  which are independent of $\alpha$. Analogous to an open TASEP, the bulk density in the HD-HD phase is $\rho_{A}=1-\beta_{A}=\rho_{B}=1-\beta_\text{eff}(N_{R})$  (current conservation). We get using Eq.~(\ref{beta-alpha-AB}) and PNC:
\begin{equation}
 \label{hd-hd-L}
\rho_\text{HD-HD}=
\begin{cases}
    (1-\beta+\mu \beta)/(1+\beta),\;\;\; \; N^{*}=L, \\
    \mu/(\mu+\beta),\;\;\; \; N^{*}=N_{0},
\end{cases}
\end{equation}
independent of $q$. Furthermore, the boundary between HD-HD and HD-DW phases is obtained from $x_{w_{2}}=1/2$ {in Eq.~(\ref{xw-hddw})},  giving $(2-\mu)\beta/(1+\beta)=q/(1+q)$ for $N^{*}=L$, and $\beta=\mu q$ for $N^{*}=N_{0}$ [see Figs.~\ref{pd}(c, f)]. 

There is another mechanism for DW formation. Depending on the parameters, upon raising $\mu$ starting from the LD-LD phase, before hitting the threshold $\alpha_\text{eff}=q/(1+q)$, it can reach another threshold satisfying $\alpha_\text{eff}=\beta_\text{eff}$, for which a DW is formed at $x=1$, beginning the LD-DW phase. Further increase in $\mu$ pushes the DW towards $x=0$, going through the  LD-DW, LD-HD (when the {DW is at} $x=1/2$; see also below) and DW-HD phases -- finally reaching the HD-HD phase.  With $\alpha_\text{eff}=\beta_\text{eff}$ in the DW phase, we get $f(N_{R})=\beta/(\alpha+\beta)$, for both $N^{*}=L,\,N_{0}$, giving DW densities as
\begin{equation}
\rho_\text{LD} = \alpha \beta/(\alpha + \beta) = 1 - \rho_\text{HD},
\label{ldw-densities-res-controlled}
\end{equation}
independent of $q$. By PNC, the DW location $x_{w_{3}}$ is
\begin{equation}
\label{xw-lddw-dwhd}
x_{w_3} = \left\{
\begin{aligned}
&\left[ \alpha + 2\beta - \mu(\alpha+\beta) - \alpha\beta \right] / \left( \alpha + \beta - 2\alpha\beta \right), \\
&\left( \alpha + \beta - \mu\alpha - \alpha\beta \right) / \left( \alpha + \beta - 2\alpha\beta \right),
\end{aligned}
\right.
\end{equation}
for $N^{*}=L$ (top) and $N^{*}=N_{0}$ (bottom). The DW shape, being controlled by $\alpha,\,\beta,\,\mu$, and $f$ is clearly {\em nonuniversal}. Substituting $x_{w_{3}}=1$ in Eq.~(\ref{xw-lddw-dwhd}) provides the boundary between LD-LD and LD-DW phases: $\mu/(1+\alpha)=\beta/(\alpha+\beta)$ for $N^{*}=L$ and $\beta=\mu$ for $N^{*}=N_{0}$ [see Figs.~\ref{pd}(a)-\ref{pd}(e)]. Next, $x_{w_{3}}=1/2$ in Eq.~(\ref{xw-lddw-dwhd}) gives the boundary between LD-DW and LD-HD phases: $\beta/(\alpha+\beta)=\mu-1/2$ for $N^{*}=L$ and $\beta/(\alpha+\beta)=1-1/2\mu$ for $N^{*}=N_{0}$ [see Figs.~\ref{pd}(b, c, e)]. Lastly, the boundary between DW-HD and HD-HD phases is obtained from $x_{w_{3}}=0$ in Eq.~(\ref{xw-lddw-dwhd}): $\alpha/(\alpha+\beta)=(2-\mu)/(1+\beta)$ for $N^{*}=L$ and $\alpha/(\alpha+\beta)=1/(\mu+\beta)$ for $N^{*}=N_{0}$.

The LD-DW and DW-HD phases are separated by the LD-HD phase line \( \mathcal{L}(\alpha,\,\beta,\,\mu) = 0 \), obtained by setting \( x_{w_3} = 1/2 \) in Eq.~(\ref{xw-lddw-dwhd}) giving
\begin{equation}
 \label{ldhd-line}
\mathcal{L}(\alpha,\,\beta,\,\mu) =
\begin{cases}
\left[(\alpha + 3\beta)\big/ 2(\alpha + \beta)\right] - \mu, \\
\left[(\alpha + \beta)\big/ 2\alpha\right] - \mu,
\end{cases}
\end{equation}
for $N^{*}=L$ (top) and $N^{*}=N_{0}$ (bottom). Stationary densities for $\alpha,\,\beta$ on (\ref{ldhd-line}) is {\em a nonuniversal} DW at $x=1/2$  but have a shape distinct from a UDW; see SM~\cite{sm}. Nonuniversal DWs are free of BLs.

A minimum current principle~\cite{astik-erwin} determines whether DWs are defect- or reservoir-induced: when the corresponding stationary currents  $J_\text{def} = q / (1 + q)^2 \;<(>)\; J_\text{res} = \alpha \beta / (\alpha + \beta) \; \big[ 1 - \alpha \beta / (\alpha + \beta) \big]
$,
DWs are controlled by the defect (reservoir). This gives the surface
\begin{equation}
 \alpha\beta/(\alpha+\beta)=q/(1+q) \label{DDW-surface}
\end{equation}
that separates the two regimes. Depending on $\mu$ and $q$, the curve~(\ref{DDW-surface}) has one or two endpoints marked\;~\protect\magpentagon\; and\;~\protect\graydiamond\; in the $\alpha$–$\beta$ plane [Figs.~\ref{pd}(a)-\ref{pd}(e), see~\cite{sm} for details]. This curve demarcates the boundary between the LD-DW and DW-LD phases, and also between DW-HD and HD-DW phases. Accordingly, DW-LD and HD-DW phases appear in the defect-controlled (green) region, while LD-DW and DW-HD lie in the reservoir-controlled (yellow) region in Fig.~\ref{pd}; see also the movies~\cite{sm}. Note that $\mathcal{L}(\alpha,\,\beta,\,\mu)$
terminates at
\begin{equation}
\label{ab-tilde}
(\tilde\alpha,\tilde\beta) =
\begin{cases}
2q/(1+q)\big(1/(2\mu-1), \; 1/(3-2\mu)\big), \\
2\mu q/(1+q)\big(1/(2\mu-1), \; 1\big),
\end{cases}
\end{equation}
for $N^{*}=L$ (top) and $N^{*}=N_{0}$ (bottom), where \((\tilde\alpha, \tilde\beta)\) is the point where  \( \mathcal{L}=0 \) and curve (\ref{DDW-surface}) intersect. The point $(\tilde\alpha, \tilde\beta)$ is a multicritical point where several phase boundaries separating the different phases meet. It is called so, since some of these phase boundaries correspond to continuous transitions with the bulk density difference being the order parameter~\cite{chaikin}. Clearly, from Eq.~\eqref{ab-tilde}, the point $(\tilde{\alpha}, \tilde{\beta})$ exists only in the range $1/2 < \mu < 3/2$ for $N^{*}=L$, and $1/2 < \mu < \infty$ for $N^{*}=N_{0}$. This point (shown as a red triangle in Fig.~\ref{pd}) shifts toward smaller $\alpha$ for large $\mu$ in the $N^{*}=N_{0}$ case and is therefore not displayed in Fig.~\ref{pd}(f). Intriguingly,  when $J_\text{def}=J_\text{res}$, i.e., when the defect ``competes'' with the reservoir, a new kind of state, {\em a  pair of DDWs}, emerges for $\mu$ satisfying the conditions for DW formation due to both the defect and reservoir. Following the logic of the DW formation caused by the defect or reservoir, we expect  one DW, say at $x=x_w^\text{def}$, in $T_\text{A}$ due to and behind the defect, and another DW at $x=x_w^\text{res}$ in $T_\text{B}$ due to and behind the reservoir. PNC gives
only a linear relation between $x_w^\text{def}$ and $x_w^\text{res}$; any pair of $(x_w^\text{def},\,x_w^\text{res})$ satisfying PNC is a valid solution. The inherent stochasticity of the underlying microscopic dynamics ensures that all such  pairs of $(x_w^\text{def},\,x_w^\text{res})$ are visited over time. This implies that a domain wall in each of $T_{A}$ and $T_{B}$ has {\em no fixed position}. This in turn gives
 one DDW in each of $T_{A}$ and $T_{B}$~\cite{niladri1}, which are just the long time averaged envelops of the moving domain walls;
see Figs.~\ref{dw-plots}(c, f). At $(\tilde\alpha, \tilde\beta)$, DDWs are fully delocalized across $T_A$ and $T_B$ [red triangular point in Figs.~\ref{pd}(b, c, e)]. Moving away from $(\tilde\alpha, \tilde\beta)$ on the curve (\ref{DDW-surface}), their spans gradually decrease, vanishing at the endpoints. Geometric constructions of the DDW profiles~\cite{sm} that depend on the parameters and $f$ (hence nonuniversal) show good agreement with the MCS results; see Fig.~\ref{deloca-along-hyperbola}. Phase transitions across (\ref{DDW-surface}) are characterized by a jump in  $\rho$, indicating a discontinuous transition; $\rho$ varies continuously across other phase boundaries. For a quick overview of the key parameters and phases, see Table~\ref{tab:key-parameters-and-phases}.
\begin{table}[t]
\centering
\renewcommand{\arraystretch}{1.05}
\caption{Key parameters and the phases}
\begin{tabular}{p{0.32\columnwidth} p{0.58\columnwidth}}
\hline
\textbf{Quantity} & \textbf{Description} \\
\hline
$\alpha,\,\beta$ & Entry and exit rates \\
$\mu=N_0/L$ & Filling factor \\
$q<1$ & Hopping rate at the defect site \\
$f(N_R)=N_R/N^{*}$ &
Reservoir feedback ($f\in[0,1]$); $N^{*}=L$ ($\mu\in[0,2]$), $N^{*}=N_0$ ($\mu\in[0,\infty)$) \\
$\alpha_{\rm eff},\,\beta_{\rm eff}$ &
Effective entry and exit rates (Eqs.~\eqref{alpha-eff}, \eqref{beta-eff}) \\
\hline
\textbf{Reservoir phases} &
LD-LD, LD-DW, LD-HD, DW-HD, HD-HD (yellow in Fig.~\ref{pd}) \\
\textbf{Defect phases} &
DW-LD, HD-DW, UDW (green in Fig.~\ref{pd}) \\
\hline
\end{tabular}
\label{tab:key-parameters-and-phases}
\end{table}

\begin{figure}[!h]
 \includegraphics[width=\columnwidth]{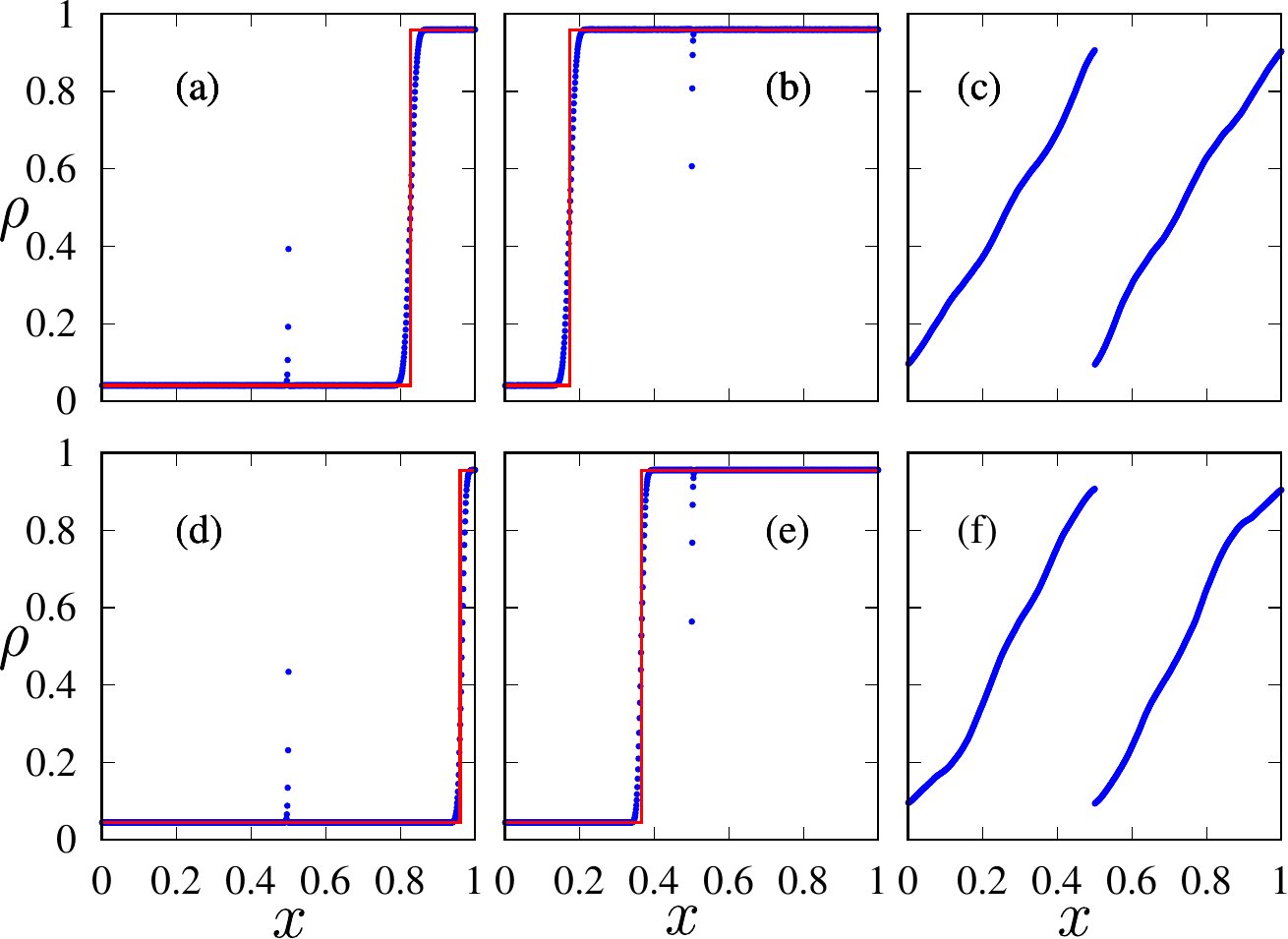}
 \caption{Reservoir-controlled LDWs in subplots~(a, b, d, e) and {a pair of completely DDWs} in~(c, f) for $q=0.1$ ($L=1000$). MFT (red solid lines) and MCS (blue circular points) results match well. Subplots~(a, b, c): $N^{*}=L$ and subplots~(d, e, f): $N^{*}=N_{0}$. (a) $\alpha=0.05$, $\beta=0.2$, $\mu=1$; (b) $\alpha=0.2$, $\beta=0.05$, $\mu=1$; (c) $\alpha=\beta=0.19$, $\mu=1$; (d) $\alpha=0.05$, $\beta=0.4$, $\mu=0.7$; (e) $\alpha=0.4$, $\beta=0.05$, $\mu=0.7$; and (f) $\alpha=0.333$, $\beta=0.1332$, $\mu=0.7$.}
 \label{dw-plots}
 \end{figure}
 
 
 \begin{figure}[!h]
 \includegraphics[width=\columnwidth]{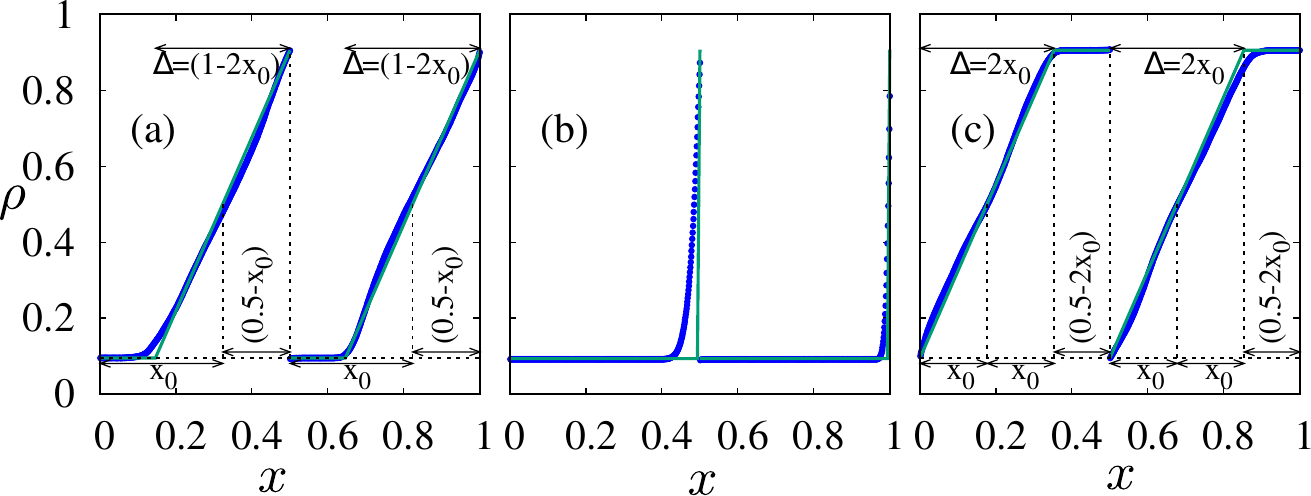}
 \caption{Partially delocalized domain walls for parameters on the curve (\ref{DDW-surface}) away from~$(\tilde\alpha,\tilde\beta)$ for $N^{*}=L$. (a) $\alpha=0.154, \beta=0.25$; (b) $\alpha=0.103, \beta=0.95$; and (c) $\alpha=0.25, \beta=0.154$ with fixed $\mu=1, q=0.1$ for all. MCS results (blue circles) match very well with geometrical constructions of the DDW profiles (green solid lines), see~\cite{sm} for the DDW position $x_{0}$ and span $\Delta$.}
 \label{deloca-along-hyperbola}
 \end{figure}

In summary, we have shown that a DW in the present model can have universal shape, being independent of all parameters except $q$, or nonuniversal, being controlled by several parameters. We expect our results should be generic: UDWs together with BLs should appear for any $f$ with monotonic dependence on the reservoir population, even though the phase diagrams should depend nonuniversally on $f$. The excellent agreement between MFT and MCS results in our work, in spite of the adhoc nature of MFT, can be qualitatively explained by the presence of the particle reservoir, which tends to weaken the density fluctuation correlation effects not included in MFT; see, e.g., related results and discussions in Refs.~\cite{tirtha-lk1,sourav-5}. Nonetheless, it will be interesting to systematically explore the effects of fluctuations on the UDWs by employing, e.g., macroscopic fluctuation theories~\cite{fluc1,fluc2}. How  the universality of the UDWs may get affected by the presence of weak nonconserving particle exchanges with the bulk~\cite{tirtha-lk1,sourav-5,tirtha-lk2} is an important theoretical question. Our basic predictions can be observed in variety of situations, e.g., ribosome density profiling experiments~\cite{riboprofile,riboden} on mRNA loops with rare ``slow codon'' or ``pause sites'' (which have distinct biological functions~\cite{cod1,cod2,cod3}) in biological cells, or simple video imaging of vehicles in closed urban roads with roadblocks and ant trails with local hindrance~\cite{antimage}
and robot swarms~\cite{robot-2024}. Localized DWs and DDWs will appear, respectively, as sharp density jumps and spatially smoothly varying profiles in such video imaging. UDWs can be identified by these images remaining unaffected even when all the parameters (except for $q$) are varied. Intriguingly, experimental detection of UDWs cannot reveal the model parameters and  functions $f$, {\em unless} BLs are also resolved.  Coarse-grained density measurement experiments do not suffice for that. Since the thickness of the BLs are of the order of a lattice size, only high resolution experiments capable of resolving such microscopic scales will detect BLs, whose heights will vary as the parameters are varied.

Experimental verification of these phenomena can be conveniently realized in model systems with  inertial active particles, such as mm- to cm-sized robots, or external field-driven spherical structures in a narrow channel where the particles cannot cross each other. While the obstruction can be created using a spatially varying driving field, the controlled entry to  (exit from) the channel from (to) the particle reservoir would require gating controlled by reservoir population. The latter and the channel population can be determined by live video imaging. 

Our model can be made more realistic by introducing inter-particle interactions and allowing for time-delays in the gates of the reservoir. From a theoretical perspective, it will be interesting to study the role of fluctuations in the DW phases by systematically going beyond MFT~\cite{fluc1,fluc2}.

{\em Acknowledgement:-} The authors thank M. Khan for helpful suggestions. A.B. thanks Alexander vol Humboldt Stiftung (Germany) for partial financial support through their research group linkage programme (2024).
 

 \clearpage
\onecolumngrid
\begin{center}
    \textbf{\large Supplemental Material for ``Universality of shocks in conserved
driven single-file motions with bottlenecks''}
\end{center}
\vspace{0.5cm}
\twocolumngrid

\section{EQUATIONS OF MOTION AND PARTICLE-HOLE SYMMETRY}
\label{eom and phs}

In this section, we present the dynamical equations of motion (EOMs) for different sites of the TASEP lane and show that particle-hole symmetry is present in our model when $N^{*} = L$, but \textit{not} when $N^{*} = N_0$. Let $\rho_j$ denote the discrete particle density at site $j$. The EOMs are as follows:
\begin{eqnarray}
\partial_t \rho_1 &=& \alpha_\text{eff} (1 - \rho_1) - \rho_1 (1 - \rho_2), \label{entry-eqn} \\
\partial_t \rho_j &=& \rho_{j-1} (1 - \rho_j) - \rho_j (1 - \rho_{j+1}), \label{bulk-eqn} \\
\partial_t \rho_L &=& \rho_{L-1} (1 - \rho_L) - \beta_\text{eff} \rho_L. \label{exit-eqn}
\end{eqnarray}
Equation~(\ref{entry-eqn}) describes the entry site ($j=1$), Eq.~(\ref{bulk-eqn}) applies to the bulk sites excluding $j = L/2$ and $j = (L/2)+1$, i.e., for $1 < j < L/2$ and $(L/2)+1 < j < L$, and Eq.~(\ref{exit-eqn}) governs the exit site ($j = L$). For the central sites $j = L/2$ and $j = (L/2)+1$, between which particles hop with reduced rate $q < 1$, the EOMs are:
\begin{eqnarray}
\partial_t \rho_{\frac{L}{2}} &=& \rho_{\frac{L}{2}-1} (1 - \rho_{\frac{L}{2}}) - q\rho_{\frac{L}{2}} (1 - \rho_{\frac{L}{2}+1}), \label{bulk-Lby2} \\
\partial_t \rho_{\frac{L}{2}+1} &=& q\rho_{\frac{L}{2}} (1 - \rho_{\frac{L}{2}+1}) - \rho_{\frac{L}{2}+1} (1 - \rho_{\frac{L}{2}+2}). \label{bulk-Lby2plus1}
\end{eqnarray}

We first consider Case~I: \( N^{*} = L \), for which the effective boundary rates are given by
\(\alpha_\text{eff} = \alpha f(N_{R})\) and \(\beta_\text{eff} = \beta \left(1 - f(N_{R})\right)\), with \(f(N_{R}) = N_{R}/L\) and \(0 \le \mu \le 2\). We aim to show that the EOMs~(\ref{entry-eqn})--(\ref{bulk-Lby2plus1}) remain invariant under the following set of transformations:
\begin{eqnarray}
\alpha &\leftrightarrow& \beta, \label{tr1} \\
\rho_j &\leftrightarrow& 1 - \rho_{L - j + 1}, \label{tr2} \\
\mu &\leftrightarrow& 2 - \mu. \label{tr3}
\end{eqnarray}
Let us first consider the entry-site EOM~(\ref{entry-eqn}). Under transformation~(\ref{tr2}), we have \( \rho_1 \leftrightarrow 1 - \rho_L \) and \( \rho_2 \leftrightarrow 1 - \rho_{L-1} \). To examine how the function \( f(N_{R}) \) transforms, we recall that \( \mu = N_{0}/L \), and using the PNC relation \( N_{0} = N_{R} + \sum_{j=1}^{L} \rho_j \), we obtain:
\begin{equation}
f(N_{R}) = \frac{N_{R}}{L} = \mu - \frac{1}{L} \sum_{j=1}^{L} \rho_j. \label{f}
\end{equation}
Applying the transformations~(\ref{tr2}) and~(\ref{tr3}) to Eq.~(\ref{f}), we get:
\begin{align}
f(N_R)
\leftrightarrow &\; (2 - \mu) - \frac{1}{L} \left( L - \sum_{j=1}^{L} \rho_j \right) \nonumber \\
=&\; 1 - \mu + \frac{1}{L} \sum_{j=1}^{L} \rho_j \nonumber \\
=&\; 1 - f(N_R). \label{f-tr}
\end{align}
Using transformations~(\ref{tr1}) and~(\ref{f-tr}) together, we get
\begin{equation}
 \alpha_\text{eff} = \alpha f(N_{R}) \leftrightarrow \beta (1-f(N_{R})) = \beta_\text{eff}.
\end{equation}
Substituting all these transformed quantities into the entry-site EOM~(\ref{entry-eqn}) maps it into the exit-site EOM~(\ref{exit-eqn}). This demonstrates that particle injection at the left boundary is equivalent to hole extraction at the right boundary. A similar invariance holds for the bulk EOMs~(\ref{bulk-eqn}) corresponding to sites \( 1 < j < L/2 \) and \( (L/2)+1 < j < L \), as well as for the defect-site EOMs~(\ref{bulk-Lby2}) and~(\ref{bulk-Lby2plus1}) at \( j = L/2 \) and \( j = (L/2) + 1 \), respectively.

We now consider Case~II: \( N^{*} = N_{0} \), for which the effective boundary rates are given by
\(\alpha_\text{eff} = \alpha f(N_{R})\) and \(\beta_\text{eff} = \beta \left(1 - f(N_{R})\right)\), where \(
f(N_{R}) = N_{R}/N_{0}\) and $0 \le \mu < \infty$. In this case, the EOMs~\eqref{bulk-eqn}, \eqref{bulk-Lby2}, and \eqref{bulk-Lby2plus1} remain invariant under the transformation~\eqref{tr2}, as for \( N^* = L \). However, the boundary-site EOMs~\eqref{entry-eqn} and \eqref{exit-eqn} are \emph{not} invariant under the transformations~\eqref{tr1}-\eqref{tr3}. Under transformation~(\ref{tr2}), the densities transform as
\(\rho_1 \leftrightarrow 1 - \rho_L\) and \(\rho_2 \leftrightarrow 1 - \rho_{L-1}\), just as in the \( N^{*} = L \) case. However, the transformation of the function \( f(N_R) \) now differs. Using the PNC, we obtain:
\begin{equation}
f(N_{R}) = \frac{N_{R}}{N_{0}} = 1 - \frac{1}{\mu L} \sum_{j=1}^{L} \rho_j. \label{f-N0}
\end{equation}
Applying transformation~(\ref{tr2}) and~(\ref{tr3}) to Eq.~(\ref{f-N0}) yields:
\begin{equation}
f(N_{R}) \leftrightarrow 1-\frac{1}{(2-\mu)}\left( 1 - \frac{1}{L}\sum_{j=1}^{L} \rho_j \right). \label{f-N0-tr}
\end{equation}
Therefore, transformations~(\ref{tr1}) and~(\ref{f-N0-tr}) together imply:
\begin{align}
\alpha_{\text{eff}} = \alpha f(N_R) \leftrightarrow \beta \left[1 - \frac{1}{(2 - \mu)} \left(1 - \frac{1}{L} \sum_{j=1}^{L} \rho_j \right) \right] \ne \beta_{\text{eff}}.
\label{abeff-N0}
\end{align}
Hence, the entry-site EOM~(\ref{entry-eqn}) does not, in general, map to the exit-site EOM~(\ref{exit-eqn}), except when \( \mu = 1 \), for which the transformation \( \alpha_{\text{eff}} \leftrightarrow \beta_{\text{eff}} \) holds—similar to the case \( N^{*} = L \). Therefore, \textit{only} for \( \mu = 1 \), the EOMs~(\ref{entry-eqn})--(\ref{bulk-Lby2plus1}) remain invariant under the transformations~(\ref{tr1})-(\ref{tr3}) in the case $N^* = N_{0}$. At \( \mu = 1 \), the phase diagram corresponding to \( N^* = N_{0} \) is identical to that of \( N^* = L \). In general, there is no transformation of $\mu$ akin to the one for \( N^* = L \), which can keep the dynamics of $N^* = N_{0}$ invariant. This shows the lack of a particle-hole symmetry in $N^* = N_{0}$.

\section{DENSITY OF THE BOUNDARY SITES IN UDW PHASE}
\label{bl-den}

In the UDW phase, the segments $T_{A}$ and $T_{B}$ are in the HD and LD phases, respectively, with bulk densities
\begin{equation}
 \label{udw-bulk-den}
 \rho_\text{LD} = \frac{q}{1 + q} = 1 - \rho_\text{HD},
\end{equation}
in both cases, $N^{*} = L$ and $N^{*} = N_{0}$. This bulk profile is accompanied by boundary layers at the sites $j = 1$ and $j = L$ whose thickness vanishes in the thermodynamic limit. We now determine the height of these boundary layers using mean-field theory.

The time evolution of the densities at the entry and exit sites, $\rho_{1}$ and $\rho_{L}$, is governed by Eqs.~(\ref{entry-eqn}) and~(\ref{exit-eqn}). In the steady state, we have $\partial_{t}\rho_{1} = \partial_{t}\rho_{L} = 0$, with $\rho_{2} = \rho_\text{HD} = 1-\rho_\text{LD}$ and $\rho_{L-1} = \rho_\text{LD}$ [see Eq.~(\ref{udw-bulk-den})] in the UDW phase. Solving these equations for $\rho_{1}$ and $\rho_{L}$ then yields:
\begin{equation}
\begin{aligned}
\rho_{1} &= \frac{\alpha_\text{eff}}{\alpha_\text{eff} + \rho_\text{LD}}, \quad
\rho_{L} = \frac{\rho_\text{LD}}{\rho_\text{LD} + \beta_\text{eff}}.
\end{aligned}
\label{rho1rhoL-udw}
\end{equation}
While $\rho_\text{LD}$ and $\rho_\text{HD}$ are independent of the normalization constant $N^{*}$ [see Eq.~(\ref{udw-bulk-den})], the effective boundary rates $\alpha_\text{eff}$ and $\beta_\text{eff}$ generally differ between the two cases $N^{*} = L$ and $N^{*} = N_{0}$. To determine them, we apply particle number conservation in each case.

\textit{Case I: $N^{*} = L$}. Here, the effective boundary rates are $\alpha_\text{eff} = \alpha N_{R}/L$ and $\beta_\text{eff} = \beta (1 - N_{R}/L)$. The number of particles in segments $T_{A}$ and $T_{B}$ are $L\rho_\text{HD}/2$ and $L\rho_\text{LD}/2$, respectively. The PNC equation becomes:
\begin{equation}
\begin{aligned}
\label{nr-L-udw}
\mu &= \frac{N_{R}}{L} + \frac{1}{L} \cdot \frac{L}{2}, \\
\Rightarrow \quad \frac{N_{R}}{L} &= \mu - \frac{1}{2}.
\end{aligned}
\end{equation}
This gives
\begin{equation}
\begin{aligned}
\alpha_\text{eff} &= \alpha\left( \mu - \frac{1}{2} \right), \quad
\beta_\text{eff} = \beta\left( \frac{3}{2} - \mu \right).
\end{aligned}
\label{eff-L-udw}
\end{equation}
Substituting Eq.~(\ref{eff-L-udw}) into Eq.~(\ref{rho1rhoL-udw}) yields:
\begin{eqnarray}
\rho_{1} &=& \left[ 1 + \frac{q}{(1 + q)\alpha\left( \mu - \frac{1}{2} \right)} \right]^{-1}, \label{rho1-L-udw} \\
\rho_{L} &=& \left[ 1 + \frac{(1 + q)\beta\left( \frac{3}{2} - \mu \right)}{q} \right]^{-1}. \label{rhoL-L-udw}
\end{eqnarray}

\textit{Case II: $N^{*} = N_{0}$}. In this case, $\alpha_\text{eff} = \alpha N_{R}/N_{0}$ and $\beta_\text{eff} = \beta(1 - N_{R}/N_{0})$. Applying PNC as before, we obtain:
\begin{equation}
 \label{nr-N0-udw}
 \frac{N_{R}}{N_{0}} = 1 - \frac{1}{2\mu},
\end{equation}
which leads to:
\begin{equation}
\begin{aligned}
\alpha_\text{eff} &= \alpha\left( 1 - \frac{1}{2\mu} \right), \quad
\beta_\text{eff} = \frac{\beta}{2\mu}.
\end{aligned}
\label{eff-N0-udw}
\end{equation}
Substituting Eq.~(\ref{eff-N0-udw}) into Eq.~(\ref{rho1rhoL-udw}) gives:
\begin{eqnarray}
\rho_{1} &=& \left[ 1 + \frac{q}{(1 + q)\alpha\left( 1 - \frac{1}{2\mu} \right)} \right]^{-1}, \label{rho1-N0-udw} \\
\rho_{L} &=& \left[ 1 + \frac{(1 + q)\beta}{2\mu q} \right]^{-1}. \label{rhoL-N0-udw}
\end{eqnarray}

Thus, while the bulk densities in the UDW phase [Eq.~(\ref{udw-bulk-den})] are universal—depending only on the parameter \( q \) and not on other control parameters \( \alpha,\,\beta,\,\mu \), or the normalization constant $N^{*}$—the boundary densities [Eqs.~(\ref{rho1-L-udw}), (\ref{rhoL-L-udw}), (\ref{rho1-N0-udw}), and (\ref{rhoL-N0-udw})] explicitly depend on all control parameters \( \alpha,\,\beta,\,\mu,\,q \) and also $N^{*}$, and are therefore nonuniversal. Furthermore, for both $N^{*} = L$ and $N^{*} = N_{0}$, as $\alpha$ and $\beta$ increase while $\mu$ and $q$ are held fixed, we observe that $\rho_{1} \to 1$ and $\rho_{L} \to 0$, whereas the domain wall remains pinned at $x = 1/2$. See Figs.~\ref{udw-univ}(a) and \ref{udw-univ}(b) for the formation of boundary layers for $N^{*}=L$ and $N^{*}=N_{0}$ respectively.

\section{DELOCALIZATION OF THE DOMAIN WALLS}
\label{deloc-of-dws}

We here discuss in details the delocalization of domain walls for $\alpha,\,\beta$ falling on the curve given by Eq.~(15) in the main text, which separates defect-controlled (green) and reservoir-controlled regions (yellow) in the $(\alpha$-$\beta)$ plane (see the phase diagrams in Fig.~2 of the main text). In the TASEP segments $T_{A}$ and $T_{B}$, nonuniversal DDWs emerge in pair along the curve~(15) in the \(\alpha\)-\(\beta\) plane (see main text), the extent of which depends explicitly on the value of \(\mu\) and $q$. Complete delocalization, i.e., when the DDWs span the entire TASEP segments, occurs at the point \((\tilde\alpha, \tilde\beta)\) in the $\alpha$-$\beta$ plane given by Eq.~(16) of the main text, where all phases meet [red triangle in Figs.~2(b), 2(c), and 2(e) of the main text]. Moving away from this point on both directions along the curve (15) of the main text, spatial spans of the DDWs across the segments of the TASEP lane gradually decreases and eventually vanishes at the end points marked\;~\protect\magpentagon\; and\;~\protect\graydiamond\ in the phase diagrams in Figs.~2(a)-2(e) of the main text. Full delocalization is seen in Fig.~3(c) (\(\alpha=\beta=0.19\), \(\mu=1\), \(q=0.1\), $N^{*}=L$) and
Fig.~3(f) (\(\alpha=0.333\), \(\beta=0.1332\), \(\mu=0.7\), \(q=0.1\), $N^{*}=N_{0}$) of the main text. In the main text, partial delocalization for $N^{*}=L$ is shown in Fig.~4(a) (\(\alpha=0.154\), \(\beta=0.25\), \(\mu=1\), \(q=0.1\)), Fig.~4(b) (\(\alpha=0.103\), \(\beta=0.95\), \(\mu=1\), \(q=0.1\)) on one side of the point \((\tilde\alpha, \tilde\beta)\), and
in Fig.~4(c) (\(\alpha=0.25\), \(\beta=0.154\), \(\mu=1\), \(q=0.1\)) on the other side. Clearly, as we move away from \((\tilde\alpha, \tilde\beta)\) on the curve (15) in the $\alpha$-$\beta$ plane, the degree of delocalization decreases, as seen in Figs.~4(a) and 4(b) of the main text. The DDWs in Figs.~4(a) and 4(c) of the main text are related by particle-hole symmetry, i.e., by the transformations~(\ref{tr1})-(\ref{tr3}).

We obtain the coordinates of the endpoints marked\;~\protect\magpentagon\; and\;~\protect\graydiamond\ of the curve~(15) of the main text separating the defect- and reservoir-controlled regions in the $\alpha$-$\beta$ plane in the following way. Point marked\;~\protect\magpentagon\; lies at the junction of the LD-LD/LD-DW, LD-LD/DW-LD, and LD-DW/DW-LD phase boundaries, whereas point marked\;~\protect\graydiamond\ is located at the intersection of the HD-HD/DW-HD, HD-HD/HD-DW, and DW-HD/HD-DW boundaries. These boundaries are  obtained for both cases \(N^{*}=L\) and \(N^{*}=N_0\), as detailed in the main text. The coordinates of these points, obtained for each case, are listed below in Table~\ref{tab-ab-coordinates}:

\begin{widetext}

\begin{table}[h!]
\centering
\small 
\renewcommand{\arraystretch}{1.8} 
\setlength{\tabcolsep}{10pt}

\caption{Coordinates of points marked\;~\protect\magpentagon\; and\;~\protect\graydiamond\ in the $\alpha$-$\beta$ plane for two representative cases}
\label{tab-ab-coordinates}

\begin{tabular}{|c|>{\centering\arraybackslash}m{6.2cm}|>{\centering\arraybackslash}m{6.2cm}|}
\hline
\diagbox{Case}{Point} & \protect\magpentagon & \protect\graydiamond \\
\hline
I. $N^{*}=L$ & $\left(\dfrac{q}{\mu(1+q)-q},\;\dfrac{q}{1+2q-\mu(1+q)}\right)$ & $\left(\dfrac{q}{\mu(1+q)-1},\;\dfrac{q}{2+q-\mu(1+q)}\right)$ \\[2ex] 
\hline
II. $N^{*}=N_{0}$ & $\left(\dfrac{\mu q}{\mu(1+q)-q},\;\mu\right)$ & $\left(\dfrac{\mu q}{\mu(1+q)-1},\;\mu q\right)$ \\
\hline
\end{tabular}
\end{table}

\end{widetext}
Hence, the coordinates of points\;~\protect\magpentagon\; and\;~\protect\graydiamond\ depend explicitly on the parameters \(\mu\) and \(q\). Since \(\alpha,\,\beta > 0\) by definition, this imposes the following conditions presented in Table~\ref{tab-ab-conditions} for the existence of points\;~\protect\magpentagon\; and\;~\protect\graydiamond\ in the two cases:

\begin{widetext}

\begin{table}[h!]
\centering
\small 
\renewcommand{\arraystretch}{1.8} 
\setlength{\tabcolsep}{10pt}

\caption{Conditions for the existence of points\;~\protect\magpentagon\; and\;~\protect\graydiamond\ for two representative cases}
\label{tab-ab-conditions}

\begin{tabular}{|c|>{\centering\arraybackslash}m{6.2cm}|>{\centering\arraybackslash}m{6.2cm}|}
\hline
\diagbox{Case}{Point} & \protect\magpentagon & \protect\graydiamond \\
\hline
I. $N^{*}=L$ & $\frac{q}{1+q} < \mu < \frac{1+2q}{1+q}$ & $\frac{1}{1+q} < \mu < \frac{2+q}{1+q}$ \\
\hline
II. $N^{*}=N_{0}$ & $\mu > \frac{q}{1+q}$ & $\mu > \frac{1}{1+q}$ \\
\hline
\end{tabular}
\end{table}

\end{widetext}

As illustrated in Fig.~2(a) of the main text, for \( N^* = L \) and \( \mu = q = 0.1 \), only the LD-LD, LD-DW, and DW-LD phases are realized, leading to the appearance of their common intersection point~\protect\magpentagon, while the HD-HD, DW-HD, and HD-DW phases are absent, and consequently, the corresponding point~\protect\graydiamond\; does not appear. In Fig.~2(b) of the main text, for \( N^* = L \) and \( \mu = 0.7,\,q = 0.1 \), the DW-HD and HD-DW phases are present, but the HD-HD phase remains absent. As a result, the point~\protect\graydiamond\; is again not observed, whereas the point~\protect\magpentagon\; persists due to the continued presence of the LD-LD, LD-DW, and DW-LD phases. In contrast, Fig.~2(c) of the main text, corresponding to \( N^* = L \) and \( \mu = 1,\,q = 0.1 \), displays both intersection points~\protect\magpentagon\; and~\protect\graydiamond, as all six phases—LD-LD, LD-DW, DW-LD, HD-HD, DW-HD, and HD-DW—are simultaneously present. A similar pattern is observed for \( N^* = N_0 \) in Figs.~2(d) and 2(e) of the main text, with \( \mu = q = 0.1 \) and \( \mu = 0.7,\,q = 0.1 \), respectively. Finally, in Fig.~2(f) of the main text, corresponding to \( \mu = 1000,\,q = 0.1 \), the phase diagram is dominated by HD-HD, HD-DW, and UDW phases, with other phases confined to very small values of \( \alpha \); hence, both intersection points~\protect\magpentagon\; and~\protect\graydiamond\; are present but not shown. These observations from Monte Carlo simulations are in full agreement with the mean-field conditions summarized in Table~\ref{tab-ab-conditions}.

 We now analyze the position and spatial span of DDWs occurring for points $(\alpha,\,\beta)$ lying on the curve (15) of the main text to determine their degree of delocalization.
Consider a pair of DDWs located with mean positions at \( x_0 \) in \( T_A \) and at \( (1/2 + x_0) \) in \( T_B \). At the delocalization transition, the conditions for a DW is satisfied in both $T_A$ and $T_B$, see Refs.\cite{blythe, niladri1, astik-erwin}. Thus the DDW envelops in $T_A$ and $T_B$ should be statistically identical. Following~\cite{niladri1,astik-erwin}, we use this to provide an analytical (geometric) construction of the DDW envelops in $T_A$ and $T_B$.   Current conservation (\( J_{\text{def}} = J_{\text{res}} \)) with $\alpha,\,\beta$ on Eq.~(15) of the main text gives
\begin{equation}
\label{ld-ddw}
\rho_{\text{LD}} = \frac{\alpha \beta}{\alpha + \beta} = \frac{q}{1 + q}=1-\rho_\text{HD},
\end{equation}
for both $N^* = L$ and $N^* = N_{0}$, see Eqs.~(6) and (12) of the main text. We apply PNC separately for \( N^* = L \) and \( N^* = N_0 \) to obtain the mean position of the DDW. Recall that the filling factor is defined as \( \mu = N_0 / L \), and for a DDW, the low-density reads $\rho_\text{LD} = \alpha_\text{eff} = 1-\rho_\text{HD}$ (according to current conservation). Hence, \( \rho_{\text{LD}} = \alpha N_R / L \) for \( N^* = L \), and \( \rho_{\text{LD}} = \alpha N_R / N_0 \) for \( N^* = N_0 \). Denoting the particle numbers in \( T_A \) and \( T_B \) as \( N_A \) and \( N_B \), respectively, PNC reads for \( N^* = L \):
\begin{align}
N_0 &= N_R + N_A + N_B, \nonumber \\
\Rightarrow \quad \mu &= \frac{\rho_\text{LD}}{\alpha}
+ \int_0^{x_0} \rho_\text{LD} \, dx
+ \int_{x_0}^{\frac{1}{2}} \rho_\text{HD} \, dx \nonumber \\
&\quad + \int_{\frac{1}{2}}^{\frac{1}{2} + x_0} \rho_\text{LD} \, dx
+ \int_{\frac{1}{2} + x_0}^{1} \rho_\text{HD} \, dx.
\label{pnc-L}
\end{align}
Solving Eq.~\eqref{pnc-L} for \( x_0 \), we obtain
\begin{equation}
x_0 = \frac{1 - \mu - \rho_\text{LD}\left( 1 - \frac{1}{\alpha} \right)}{2(1 - 2\rho_\text{LD})},
\label{x0-L}
\end{equation}
where \( \rho_\text{LD} \) is given by Eq.~\eqref{ld-ddw}. Similarly, for \( N^* = N_0 \), we get
\begin{equation}
x_0 = \frac{1 - \mu - \rho_\text{LD}\left( 1 - \frac{\mu}{\alpha} \right)}{2(1 - 2\rho_\text{LD})},
\label{x0-N0}
\end{equation}
with \( \rho_\text{LD} \) again given by Eq.~\eqref{ld-ddw}. Depending on $\alpha,\,\beta$ for fixed values of $\mu$ and $q$, the DW position, $x_{0}$, obtained in Eqs.~(\ref{x0-L}) and (\ref{x0-N0}) can be greater than [see Figs.~4(a) and 4(b) of the main text], equal to [see Figs.~3(c) and 3(f) of the main text], or less than [see Fig.~4(c) of the main text] 1/4. DDW span, $\Delta$, is obtained from the geometric constructions of DDWs in Fig.~4 of the main text as
\begin{equation}
 \label{DDW-span}
\Delta =
\begin{cases}
    1-2x_{0}, \;\;\; \text{for} \; x_{0}>1/4, \\
    1/2, \;\;\; \text{for} \; x_{0}=1/4, \\
    2x_{0}, \;\;\; \text{for} \; x_{0}<1/4,
\end{cases}
\end{equation}
for both $N^{*}=L$ and $N^{*}=N_{0}$.

We will now compare $x_{0}$ and $\Delta$ obtained from MCS in Fig.~4 of the main text with those obtained analytically in Eqs.~(\ref{x0-L}), (\ref{x0-N0}), and (\ref{DDW-span}) above. We consider only the case $N^{*}=L$ with $\mu=1,q=0.1$ for which the point of complete delocalization is $(\tilde{\alpha}, \tilde{\beta})=(0.19,0.19)$, see Eq.~(16) of the main paper. In Fig.~4(a) of the main text with \( \alpha = 0.154 \), \( \beta = 0.25 \), we get \( x_0 = 0.32 \) and \( \Delta = 1 - 2x_0 = 0.36 \). In Fig.~4(b) of the main text, with \( \alpha = 0.103 \), \( \beta = 0.95 \), we find \( x_0 = 0.49 \) and \( \Delta = 1 - 2x_0 = 0.02 \). These results indicate that the degree of delocalization of the DW decreases as one moves away from the point of complete delocalization, i.e., $(\tilde{\alpha}, \tilde{\beta})$. Finally, in Fig.~4(c) of the main text, for \( \alpha = 0.25 \), \( \beta = 0.154 \), we obtain \( x_0 = 0.18 \) and \( \Delta = 2x_0 = 0.36 \). All these analytical results match very well with the values of $x_{0}$ and $\Delta$ obtained by MCS.

\section{DOMAIN WALLS IN UDW AND LD-HD PHASES}
\label{dw in udw and ldhd}

We present in this section the domain wall profiles for representative parameter sets lying in the UDW and LD-HD phases.

In the defect-dominated UDW phase, the domain wall is pinned at \( x = 1/2 \) irrespective of any control parameters, separating HD and LD regions appearing on segments \( T_A \) and \( T_B \), respectively. Boundary layers of vanishing thickness in TL form on both sides of the reservoir at sites \( j = 1 \) and \( j = L \). The bulk densities on either domains of the UDW are given by
\begin{equation}
 \rho_\text{LD} = \frac{q}{1 + q} = 1-\rho_\text{HD}, \label{udw-den}
\end{equation}
indicating that both densities depend solely on the defect strength \( q \), and are independent of \( \alpha \), \( \beta \), \( \mu \), or the choice of the function \( f \). The resulting domain wall is thus universal. Figs.~\ref{udw-univ}(a) (\( N^{*} = L \)) and \ref{udw-univ}(b) (\( N^{*} = N_0 \)) illustrate such universal domain walls for two distinct parameter sets with identical \( q = 0.1 \). In both cases, the wall position \( x_w = 1/2 \) and height  given by
\[\rho_\text{HD}-\rho_\text{LD} = \frac{1-q}{1+q}=0.816,
\]
remain independent of $\alpha,\,\beta,\,\mu,\,f$, demonstrating the universal nature of UDWs vis-\'a-vis these parameters; see Fig.~\ref{udw-univ}. Furthermore, boundary layers are observed at both ends of the TASEP lane; see Fig.~\ref{udw-univ}. This is consistent with MFT predictions.

\begin{figure*}[htb]
  \includegraphics[width=\textwidth]{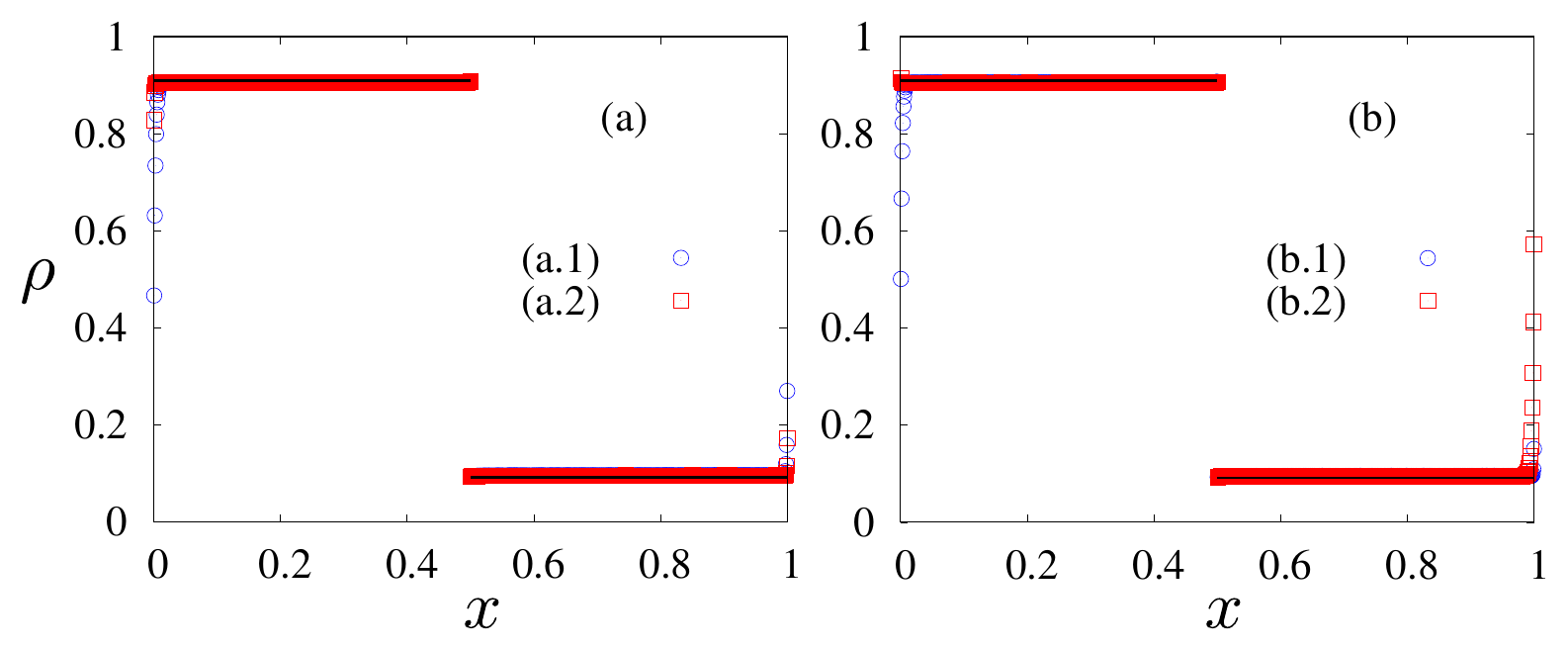}
  \caption{Universal domain walls in the UDW phase. (a) corresponds to \( N^{*} = L \), with parameters: (a.1) $\alpha=0.8,\,\beta=0.4,\,\mu=0.7,\,q=0.1$ and (a.2) $\alpha=\beta=1,\,\mu=1,\,q=0.1$; while (b) corresponds to \( N^{*} = N_0 \), with parameters: (b.1) $\alpha=0.6,\,\beta=0.8,\,\mu=0.7,\,q=0.1$ and (b.2) $\alpha=150,\,\beta=300,\,\mu=1000,\,q=0.1$. The domain walls in both cases occur at the same position, \( x_w = 1/2 \), and have the same height \( \frac{1-q}{1 + q} = 0.816 \), which depends only on the defect strength \( q \), and not on \( \alpha \), \( \beta \), \( \mu \), \( N^{*} \) or \( f \). This demonstrates the universality of the UDW profile with respect to these parameters. MFT predictions (solid black line) and MCS results (colored points) match very well. The UDWs are accompanied by boundary layers at the entry- and exit-ends of TASEP.}
  \label{udw-univ}
\end{figure*}

\begin{figure*}[htb]
  \includegraphics[width=\textwidth]{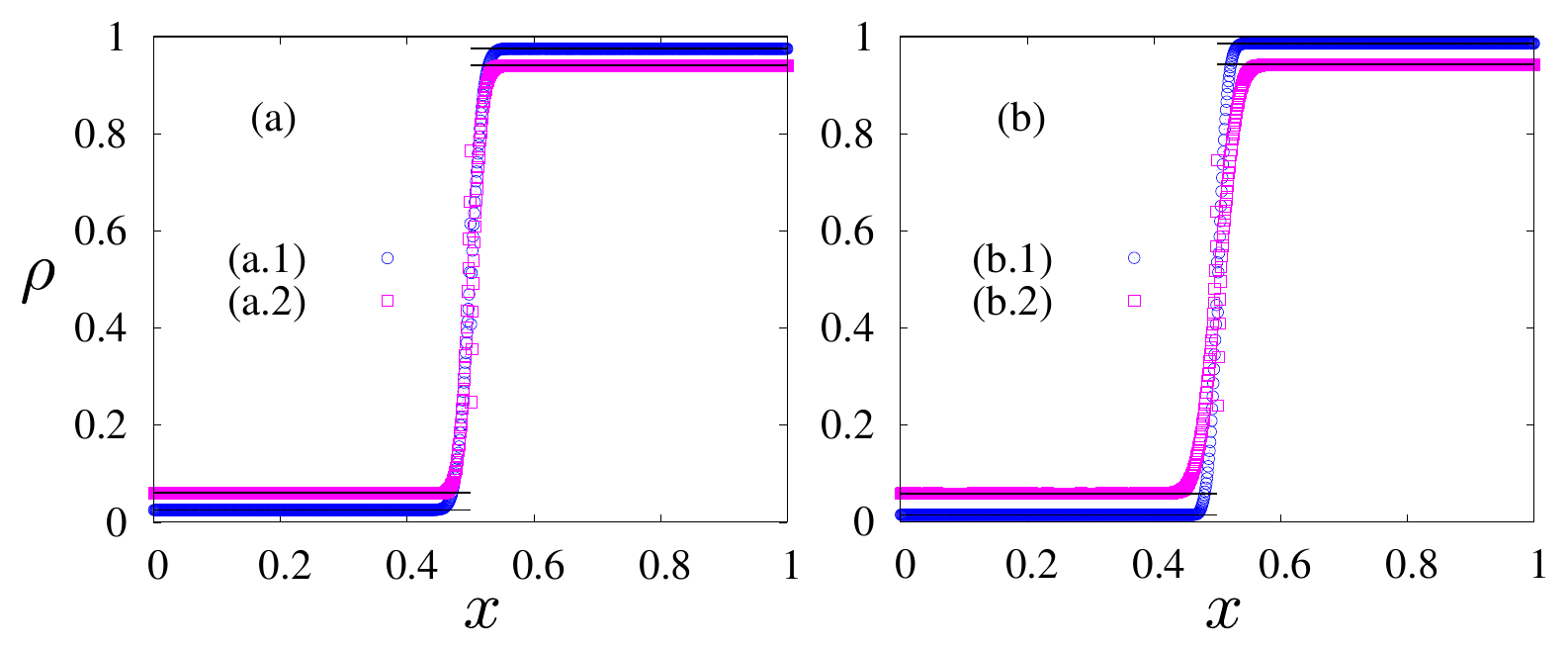}
  \caption{Nonuniversal domain walls in the LD-HD phase. (a): \( N^{*} = L \) with (a.1) $\alpha=\beta=0.05,\,\mu=1,\,q=0.1$ and (a.2) $\alpha=0.3,\,\beta=0.075,\,\mu=0.7,\,q=0.1$; (b): \( N^{*} = N_0 \) with (b.1) $\alpha=0.05,\,\beta=0.02,\,\mu=0.7,\,q=0.1$ and (b.2) $\alpha=0.07,\,\beta=0.35,\,\mu=3,\,q=0.1$. In all cases, domain walls are located at \( x = 1/2 \), but their height, \( 1 - \frac{2\alpha\beta}{\alpha + \beta} \), depends on the boundary parameters \( \alpha \) and \( \beta \), reflecting their nonuniversal nature. MFT predictions (solid black lines) show excellent agreement with MCS data (colored points).}
  \label{ldhd_nonuniv}
\end{figure*}

In addition to the universal domain walls observed in the UDW phase, we also find domain walls along the reservoir-controlled LD-HD phase line \( \mathcal{L}(\alpha,\,\beta,\,\mu) = 0 \) (see Eq.~(14) in the main text) with bulk densities in the LD and HD domain appearing on the segments $T_{A}$ and $T_{B}$ respectively given by
\begin{equation}
 \rho_\text{LD} = \frac{\alpha\beta}{\alpha+\beta} = 1-\rho_\text{HD}, \label{ldhd-den}
\end{equation}
These walls are likewise located at \( x = 1/2 \), similar to the UDWs, but their height
\[\rho_\text{HD}-\rho_\text{LD} =
1 - \frac{2\alpha\beta}{\alpha + \beta},
\]
depends explicitly on the boundary parameters \( \alpha \) and \( \beta \), rendering them nonuniversal. Apart from the issue of universality, although both UDWs and nonuniversal DWs are located at $x=1/2$, there is a fundamental distinction between the two. The HD segment of a UDW covers $T_A$ and the LD segment covers $T_B$. In contrast, a nonuversal DW formed for $\alpha,\,\beta$ on the LD-HD line in the $\alpha$-$\beta$ plane has its LD segment covering $T_A$ and the HD segment covering $T_B$. The mean-field theory predictions for the density profiles in the LD-HD phase are validated by Monte Carlo simulation results, as shown in Figs.~\ref{ldhd_nonuniv}(a) for \(N^{*} = L\) and \ref{ldhd_nonuniv}(b) for \(N^{*} = N_0\).


\begin{thebibliography}{99}
 \bibitem{cell} B. Alberts, A. Johnson, P. Walter, and J. Lewis, Molecular Biology of the Cell, 5th ed.(Garland Publishing Inc., New York, 2008).

\bibitem{mrna22} S. E. Wells, E. Hillner, R. D. Vale and A. B. Sachs, Circularization of mRNA by Eukaryotic
Translation Initiation Factors, Mol. Cell. {\bf 2}, 135 (1998).

\bibitem{mrna33} S. Wang, K. S. Browning and W.
A. Miller, A viral sequence in the 3'‐untranslated region mimics a 5' cap in facilitating translation of uncapped mRNA, EMBO J. {\bf 16}, 4107 (1997).

\bibitem{mrna44} Y. Nakamura, T. Gojobori, and T. Ikemura, Codon usage tabulated from international DNA sequence databases: status for the year 2000, Nucleic Acids Res. {\bf 28}, 292
(2000).

\bibitem{traffic1} D. Chowdhury, L. Santen and A. Schadschneider, Statistical physics of vehicular traffic and some related systems, Phys. Rep. {\bf 329}, 199 (2000).

\bibitem{traffic2} D. Helbing, Traffic and related self-driven many-particle systems, Rev. Mod. Phys. {\bf 73}, 1067 (2001).

\bibitem{bus-route} N. P. N. Ngoc, H. A. Thi, and N. V. Vinh, Exactly solvable dual bus-route model, Phys. Rev. E {\bf 110}, 054130 (2024).

\bibitem{ant1} D. Chowdhury, V. Guttal, K. Nishinari and A. Schadschneider, A cellular-automata model of flow in ant-trails: non-monotonic variation of speed with density, arXiv:cond-mat/0201207v3, J. Phys.A {\bf 35}, L573 (2002).

\bibitem{ant2} N. P. N. Ngoc, H. A. Thi, and N. V. Vinh, An exactly solvable model for single-lane unidirectional ant traffic, Physica A: Statistical Mechanics and its Applications {\bf 651},  130022 (2024).

\bibitem{robot-2024} L. Alonso-Llanes ,A. Garcimartín,  and I. Zuriguel, Phys. Rev. Research {\bf 6}, L022037 (2024).

\bibitem{krug} J. Krug, Boundary-induced phase transitions in driven diffusive systems, Phys. Rev. Lett. {\bf 67}, 1882 (1991).

\bibitem{derrida1} B. Derrida, S. A. Janowsky, J. L. Lebowitz, and E. R. Speer, Exact solution of the totally asymmetric simple exclusion process: Shock profiles, J. Stat. Phys. {\bf 73}, 813 (1993).

\bibitem{derrida2} B. Derrida and M. R Evans, Exact correlation functions in an asymmetric exclusion model with open boundaries, Journal de Physique I {\bf 3}, 311 (1993).

\bibitem{macdonald} C. T. MacDonald, J. H. Gibbs, and A. C. Pipkin, Kinetics of biopolymerization on nucleic acid templates,  Biopolymers {\bf 6}, 1 (1968).

\bibitem{derrida3} B. Derrida, M. R. Evans, V. Hakim, and V. Pasquier, Exact solution of a 1d asymmetric exclusion model using a matrix formulation, J. Phys. A: Math. and Gen. {\bf 26}, 1493–1517 (1993).

\bibitem{kolomeisky-1998} A. B. Kolomeisky, Asymmetric simple exclusion model with local inhomogeneity, J. Phys. A: Math. Gen. {\bf 31}, 1153 (1998).

\bibitem{blythe} R. A. Blythe and M. R. Evans, Nonequilibrium steady states of matrix-product form: a solver's guide, J. Phys. A {\bf 40}, R333 (2007).

\bibitem{chou} T. Chou, K. Mallick, and R. K. P. Zia, Non-equilibrium statistical mechanics: from a paradigmatic model to biological transport, Rep. Prog. Phys. {\bf 74}, 116601 (2011).

\bibitem{atri1} A. Goswami, M. Chatterjee, and S. Mukherjee, Steady states and phase transitions in heterogeneous asymmetric exclusion processes, J. Stat. Mech. (2022) 123209.

\bibitem{atri2} A. Goswami, U. Dey, S. Mukherjee, Nonequilibrium steady states in coupled asymmetric and symmetric exclusion processes, Phys. Rev. E {\bf 108}, 054122 (2023).

\bibitem{sm-ab-tasep} S. Mukherjee and A. Basu, Nonuniform asymmetric exclusion process: Stationary densities and domain walls, arXiv: 2410.22516.

\bibitem{lebo} S. A. Janowsky and J. L. Lebowitz, Finite-size effects and shock fluctuations in the asymmetric simple exclusion process, Phys. Rev. A {\bf 45}, 618 (1992).

\bibitem{lebo1}{ S. A. Janowsky and J. L. Lebowitz, Exact results for the asymmetric simple exclusion process with a blockage, J. Stat. Phys. {\bf 77}, 35 (1994).}

\bibitem{mustansir} G. Tripathy and M. Barma, Driven lattice gases with quenched disorder: Exact results and different macroscopic regimes, Phys. Rev. E {\bf 58}, 1911 (1998).

\bibitem{ha-nijs-2002} M. Ha and M. den Nijs, Macroscopic car condensation in a parking garage, Phys. Rev. E {\bf 66}, 036118 (2002).

\bibitem{erwin-defect} P. Pierobon, M. Mobilia, R. Kouyos, and E. Frey, Bottleneck-induced transitions in a minimal model for intracellular transport, Phys. Rev. E {\bf 74}, 031906 (2006).

\bibitem{hinsch} H. Hinsch and E. Frey, Bulk-driven nonequilibrium phase transitions in a mesoscopic ring, Phys. Rev. Lett. {\bf 97}, 095701 (2006).

\bibitem{basu-mohanty} U. Basu and P. K. Mohanty, Spatial correlations in exclusion models corresponding to the zero-range process, J. Stat. Mech., online at:  stacks.iop.org/JSTAT/2010/L03006, doi 10.1088/1742-5468/2010/03/L03006

\bibitem{corstin-2012} O. Costin, J. L. Lebowitz, E. R. Speer and A. Troiani, The blockage problem, Bull. Inst. Math. Acad. Sin. N. S. {\bf 8}, 49 (2013), arXiv:1207.6555.

\bibitem{rakesh1}  R. Chatterjee, A. K. Chandra, and A. Basu, Phase transition and phase coexistence in coupled rings with driven exclusion processes, Phys. Rev. E {\bf 87}, 032157 (2013)

\bibitem{niladri1} N. Sarkar and A. Basu, Nonequilibrium steady states in asymmetric exclusion processes on a ring with bottlenecks, Phys. Rev. E {\bf 90}, 022109 (2014).

\bibitem{r-basu} R Basu, V Sidoravicius and A Sly, Last Passage Percolation with a Defect Line and the Solution of the Slow Bond Problem, arXiv:1408.3464.

\bibitem{niladri-tirtha} T. Banerjee, N. Sarkar and A. Basu, Generic nonequilibrium steady states in an exclusion process on an inhomogeneous ring, J. Stat. Mech. P01024 (2015).

\bibitem{rakesh2} R. Chatterjee, A. K. Chandra, and A. Basu, Asymmetric exclusion processes on a closed network with bottlenecks, J. Stat. Mech. (2015) P01012

\bibitem{soh} H. Soh, Y. Beck, M. Ha and H. Jeong,  Effects of a local defect on one-dimensional nonlinear surface growth, Phys. Rev. E {\bf 95}, 042123 (2017).

\bibitem{tirtha-qxtasep} T. Banerjee and A. Basu, Smooth or shock: Universality in closed inhomogeneous driven single file motions, Phys. Rev. Research {\bf 2}, 013025 (2020).

\bibitem{parna-anjan} P. Roy, A.K. Chandra, and A. Basu, Pinned or moving: states of a single shock in a ring, Phys. Rev. E {\bf 102}, 012105 (2020).

\bibitem{atri3} A. Goswami, R. Chatterjee, and S. Mukherjee, Defect versus defect: stationary states of single file marching in periodic landscapes with road blocks, arXiv: 2402.08499.

\bibitem{motor} M. Schliwa and G. Woehlke, Molecular motors, Nature \textbf{422}, 759 (2003).

\bibitem{tripathy-barma97} G. Tripathy and M. Barma, Steady state and dynamics of driven diffusive systems with quenched disorder, Phys. Rev. Lett. \textbf{78}, 3039 (1997).

\bibitem{bhatia23} N. Bhatia and A. K. Gupta, Role of site-wise dynamic defects in a resource-constrained exclusion process, Chaos, Solitons \& Fractals \textbf{167}, 113109 (2023).

\bibitem{waclaw19} B. Waclaw, J. Cholewa-Waclaw, and P. Greulich, Totally asymmetric exclusion process with site-wise dynamic disorder, J. Phys. A: Math. Theor. \textbf{52}, 065002 (2019).

\bibitem{ha-timonen-nijs} M. Ha, J. Timonen and M. den Nijs, Queuing transitions in the asymmetric simple exclusion process, Phys. Rev. E {\bf 68}, 056122 (2003).


\bibitem{schmidt15} J.~Schmidt, V.~Popkov, and A.~Schadschneider, ``Defect-induced phase transition in the asymmetric simple exclusion process,'' \textit{Europhys.\ Lett.} \textbf{110}, 20008 (2015).

\bibitem{soh18} H.~Soh, M.~Ha, and H.~Jeong,
``Jamming and condensation in one-dimensional driven flow,'' \textit{Phys.\ Rev.\ E} \textbf{97}, 032120 (2018).


\bibitem{reser1} D. A. Adams, B. Schmittmann, and R. K. P. Zia, Far from equilibrium transport with constrained resources, J. Stat. Mech. (2008) P06009.

\bibitem{reser2}  L. Jonathan Cook and R. K. P. Zia, Feedback and fluctuations in a totally asymmetric simple exclusion process with finite resources, J. Stat. Mech. (2009) P02012.

\bibitem{reser3} L. J. Cook, R. K. P. Zia, and B. Schmittmann, Competition between multiple totally asymmetric simple exclusion processes for a finite pool of resources, Phys. Rev. E {\bf 80}, 031142 (2009).

\bibitem{klumpp1} R. Lipowsky, S. Klumpp, and T. M. Nieuwenhuizen, Random walks of cytoskeletal motors in open and closed compartments, Phys. Rev. Lett. {\bf 87}, 108101 (2001).

\bibitem{klumpp2} S. Klumpp and R. Lipowsky, Traffic of molecular motors through tube-like compartments, J. Stat. Phys. {\bf 113}, 233 (2003).

\bibitem{klumpp3} S. Klumpp and R. Lipowsky, Asymmetric simple exclusion processes with diffusive bottlenecks, Phys. Rev. E {\bf 70}, 066104 (2004).

\bibitem{ciandrini} L. D. Fernandes and L. Ciandrini, Driven transport on a flexible polymer with particle recycling: A model inspired by transcription and translation, Phys. Rev. E {\bf 99}, 052409 (2019).

\bibitem{dauloudet} O. Dauloudet, I. Neri, J.-C. Walter, J. Dorignac, F. Geniet, and A. Parmeggiani, Modelling the effect of ribosome mobility on the rate of protein synthesis, Eur. Phys. J. E {\bf 44}, 1 (2021).

\bibitem{astik-parna} A. Haldar, P. Roy, and A. Basu, Asymmetric exclusion processes with fixed resources: Reservoir crowding and steady states, Phys. Rev. E {\bf 104}, 034106 (2021).

\bibitem{sourav1} S. Pal, P. Roy, and A. Basu, Availability, storage capacity, and diffusion: Stationary states of an asymmetric exclusion process
connected to two reservoirs, Phys. Rev. E {\bf 110}, 054104 (2024).

\bibitem{sourav2} S. Pal, P. Roy, and A. Basu, Distributed fixed resources exchanging particles: Phases of an asymmetric exclusion process connected to two reservoirs, Phys. Rev. E {\bf 111}, 034109 (2025).

\bibitem{sourav3} S. Pal, P. Roy, and A. Basu, Stationary densities and delocalized domain walls in asymmetric exclusion processes competing for finite pools of resources, arXiv:2509.23983.

\bibitem{astik-erwin} A. Haldar, P. Roy, E. Frey, and A. Basu, Availability versus carrying capacity: Phases of asymmetric exclusion processes competing for finite pools of resources, Phys. Rev. E {\bf 111}, 014154 (2025).

\bibitem{seppa} T. Sepp\"al\"ainen, Existence of hydrodynamics for the totally asymmetric simple K-exclusion process, Ann. Probab. {\bf 27}, 361
(1999).

\bibitem{supply-demand} C. A. Brackley, M.C. Romano and M. Thiel, The Dynamics of Supply and Demand in mRNA
Translation, PLoS Computational Biology {\bf 7}, e1002203 (2011).

\bibitem{sm} Supplemental Material containing details on the particle-hole symmetry for Case I, additional calculational details and movies.

\bibitem{kolomeisky-TASEP-phase} A. B. Kolomeisky, Phase diagram of one-dimensional driven lattice
gases with open boundaries, J. Phys. A: Math. Gen. {\bf 31}, 6911 (1998).

\bibitem{kavcic2025} B.~Kavčič and G.~Tkačik,
``Token-driven totally asymmetric simple exclusion processes,'' Phys. Rev. E \textbf{111}, 054122 (2025).

\bibitem{jindal-arvind} B. Pal and A. K. Gupta,  ``Exclusion process on two intersecting lanes with constrained resources: Symmetry breaking and shock dynamics'', Phys. Rev. E  {\bf 104}, 014138 (2021).

\bibitem{arvind1} B. Pal and A. K. Gupta,  ``Reservoir crowding in a resource-constrained exclusion process with a dynamic defect'', Phys. Rev. E  {\bf 106}, 044130 (2022).

\bibitem{chaikin} P. M. Chaikin and T. C. Lubensky, Principles of Condensed Matter Physics (Cambridge University Press, Cambridge, 2000).

\bibitem{brackley} C. A. Brackley, M. C. Romano, and M. Thiel, Slow sites in an exclusion process with limited resources, Phys. Rev. E {\bf 82}, 051920 (2010).


\bibitem{cook-slow} L. J. Cook, J. J. Dong, and A. LaFleur, Interplay between finite resources and a local defect in an asymmetric simple exclusion process, Phys. Rev. E {\bf 88}, 042127 (2013).

\bibitem{tirtha-lk1} T. Banerjee, A. K. Chandra, and A. Basu, Phase coexistence and particle nonconservation in a closed asymmetric exclusion process with inhomogeneities, Phys. Rev. E {\bf 92}, 022121 (2015).

\bibitem{sourav-5} S. Pal and A. Basu, Stationary densities in a weakly nonconserving asymmetric exclusion processes with finite resources, arXiv:2602.08405. 



\bibitem{fluc1} L. Bertini, A. D. Sole, D. Gabrielli, G. Jona-Lasinio and C. Landim, Macroscopic fluctuation theory, Rev. Mod. Phys. {\bf 87}, 593 (2015).

\bibitem{fluc2} B. Doyon, G. Perfetto, T. Sasamoto, and T. Yoshimura, Ballistic macroscopic fluctuation theory, SciPost Phys. {\bf 15}, 136 (2023).

\bibitem{tirtha-lk2} B. Daga, S. Mondal, A. K. Chandra, T. Banerjee, and A. Basu, Nonequilibrium steady states in a closed inhomogeneous asymmetric exclusion process with generic particle nonconservation, Phys. Rev. E {\bf 95}, 012113 (2017).

\bibitem{riboprofile} N. T. Ingolia et al., Genome-Wide Analysis in Vivo of Translation with Nucleotide Resolution Using Ribosome Profiling, Science {\bf 324}, 218 (2009).

\bibitem{riboden} Y. Arava,  F. Edward Boas, P. O. Brown and D. Herschlag, Dissecting eukaryotic translation and its control by ribosome density mapping, Nucleic Acids Res. {\bf 33}, 2421 (2005).

\bibitem{cod1} R. Brockmann, A. Beyer, J. J. Heinisch, and T. Wilhelm, Posttranscriptional Expression Regulation: What Determines Translation Rates?, PLOS Comput. Biol. {\bf 3}, e57 (2007).

\bibitem{cod2} I. J. Purvis, A. J. E. Bettany, T. C. Santiago, J. R. Coggins, K. Duncan, R. Eason, and A. J. P. Brown, The efficiency of folding of some proteins is increased by controlled rates of translation in vivo: A hypothesis, J. Mol. Biol. {\bf 193}, 413
(1987).

\bibitem{cod3} M. C. Romano, M. Thiel, I. Stansfield, and C. Grebogi, Queueing Phase Transition: Theory of Translation, Phys. Rev. Lett. {\bf 102}, 198104 (2009).

\bibitem{antimage} J. A. Sabattini, F. Sturniolo, M. Bollazzi, L. A. Bugnon, AntTracker: A low-cost and efficient computer vision approach to research leaf-cutter ants behavior, Smart Agricultural Technology {\bf 5}, 100252 (2023).


\end{thebibliography}
\end{document}